\documentclass[aps, prb, reprint, showpacs, showkeys, twocolumn, superscriptaddress, notitlepage]{revtex4-1}
\usepackage{graphicx} 
\usepackage{braket}
\usepackage{color}
\usepackage{times}
\usepackage{amsmath}
\usepackage{lipsum}
\usepackage{multirow}  
\usepackage{adjustbox}  
\usepackage{textcomp} 

\DeclareTextSymbol{\degre}{T1}{6}
\DeclareTextSymbol{\degre}{OT1}{23}

\begin{document}

\title{An artificial Rb atom in a semiconductor with lifetime-limited linewidth}

\author{Jan-Philipp Jahn}
\affiliation{Department of Physics, University of Basel, Klingelbergstrasse 82, CH-4056 Basel, Switzerland}

\author{Mathieu Munsch}
\affiliation{Department of Physics, University of Basel, Klingelbergstrasse 82, CH-4056 Basel, Switzerland}

\author{Lucas B\'eguin}
\affiliation{Department of Physics, University of Basel, Klingelbergstrasse 82, CH-4056 Basel, Switzerland}

\author{Andreas V. Kuhlmann}
\affiliation{Department of Physics, University of Basel, Klingelbergstrasse 82, CH-4056 Basel, Switzerland}

\author{Martina Renggli}
\affiliation{Department of Physics, University of Basel, Klingelbergstrasse 82, CH-4056 Basel, Switzerland}

\author{Yongheng Huo}
\affiliation{Institute of Semiconductor and Solid State Physics, Johannes Kepler University Linz, Altenbergerstrasse 69, A-4040 Linz, Austria}

\author{Fei Ding}
\affiliation{Institute for Integrative Nanosciences, IFW Dresden, Helmholtzstrasse 20, D-01069 Dresden, Germany}

\author{Rinaldo Trotta}
\affiliation{Institute of Semiconductor and Solid State Physics, Johannes Kepler University Linz, Altenbergerstrasse 69, A-4040 Linz, Austria}

\author{Marcus Reindl}
\affiliation{Institute of Semiconductor and Solid State Physics, Johannes Kepler University Linz, Altenbergerstrasse 69, A-4040 Linz, Austria}

\author{Oliver G. Schmidt}
\affiliation{Institute for Integrative Nanosciences, IFW Dresden, Helmholtzstrasse 20, D-01069 Dresden, Germany}

\author{Armando Rastelli}
\affiliation{Institute of Semiconductor and Solid State Physics, Johannes Kepler University Linz, Altenbergerstrasse 69, A-4040 Linz, Austria}

\author{Philipp Treutlein}
\affiliation{Department of Physics, University of Basel, Klingelbergstrasse 82, CH-4056 Basel, Switzerland}

\author{Richard J. Warburton}
\affiliation{Department of Physics, University of Basel, Klingelbergstrasse 82, CH-4056 Basel, Switzerland}

\date{\today}

\begin{abstract}
We report results important for the creation of a best-of-both-worlds quantum hybrid system consisting of a solid-state source of single photons and an atomic ensemble as quantum memory. We generate single photons from a GaAs quantum dot (QD) frequency-matched to the Rb D2-transitions and then use the Rb transitions to analyze spectrally the quantum dot photons. We demonstrate lifetime-limited QD linewidths (1.42 GHz) with both resonant and non-resonant excitation. The QD resonance fluorescence in the low power regime is dominated by Rayleigh scattering, a route to match quantum dot and Rb atom linewidths and to shape the temporal wave packet of the QD photons. Noise in the solid-state environment is relatively benign: there is a blinking of the resonance fluorescence at MHz rates but negligible dephasing of the QD excitonic transition. We therefore demonstrate significant progress towards the realization of an ideal solid-state source of single photons at a key wavelength for quantum technologies.
\end{abstract}

\maketitle

\section{Introduction}
Establishing the hardware for a quantum network is a challenging task. A source of indistinguishable single photons is required along with a means to store the single photons at each node. Single semiconductor quantum dots are excellent sources of single photons: they are bright, robust and fast emitters \cite{Shields2007,Warburton2013}. A single quantum dot mimics a two-level atom closely such that single photons can be generated either by spontaneous emission from the upper level \cite{He2013} or by coherent scattering of a resonant laser \cite{Nguyen2011,Matthiesen2012,Matthiesen2013}. Subsequently emitted photons are close to indistinguishable \cite{Santori2002}. However, achieving the lifetime-limit has been an elusive goal \cite{Hogele2004,Kuhlmann2013a}, and the wavelength coverage is limited. 

Independently, atomic ensembles have developed into one of the best platforms for optical quantum memories~\cite{Lvovsky2009,Bussieres2013}. The combination of strong absorption and long ground state hyperfine coherence has allowed storage times of miliseconds and efficiencies higher than 75\,\% to be achieved in these systems~\cite{Radnaev2010,Bao2012,Chen2013,Bimbard2014}. Moreover, schemes for broadband operation with single photons at the GHz level have been proposed~\cite{Rakher2013} and also demonstrated experimentally~\cite{Michelberger2015}; single photons emitted by a single atom were stored in a Bose-Einstein condensate of the same species and used to produce entanglement between the two remote systems~\cite{Lettner2011}. 

A semiconductor-cold atom quantum hybrid would combine the advantage of the semiconductor (straightforward single photon generation, large oscillator strength) with the advantage of the cold atoms (slow decoherence) whilst avoiding the disadvantage of the semiconductor (fast decoherence \cite{Warburton2013}) and the disadvantage of the cold atoms (complex single photon generation \cite{Darquie2005}). This would constitute an implementation of a quantum repeater using single photon sources and memories \cite{Sangouard2007}. Unfortunately, the workhorse systems are mismatched in frequency: self-assembled InGaAs quantum dots emit typically around 950 nm; the D1 and D2 transitions of the Rb atoms lie at 795 and 780 nm. We note that a frequency match has been achieved with Cs~\cite{Ulrich2014}; a link has also been established with a transition of the Yb$^{+}$ ion~\cite{Meyer2015}; a trapped molecule produces single photons at the Na frequency~\cite{Siyushev2014}; and a new quantum dot growth procedure has led to a first hybrid experiment with Rb~\cite{Akopian2011}.  A high quality semiconductor source of single photons frequency-matched to the Rb transitions is highly desirable. 

We present here a close-to-ideal semiconductor source of single photons at the Rb D2 wavelength. The emission frequency can be tuned through all the D2-hyperfine lines. We demonstrate lifetime-limited quantum dot linewidths. This points to negligible upper level dephasing and allows us to create photons by coherent Rayleigh scattering with weak, resonant excitation. We find that all our experiments (spectral analysis, intensity autocorrelation, decay dynamics) can be described in terms of a two-level atom with a common set of parameters. The only significant source of noise is slow relative to radiative emission and results in a telegraph-like blinking behavior. Apart from this the system behaves in an ideal way despite the complexity of the solid-state environment. 

\section{Sample}
\begin{figure}[tb]
\includegraphics[width=85mm]{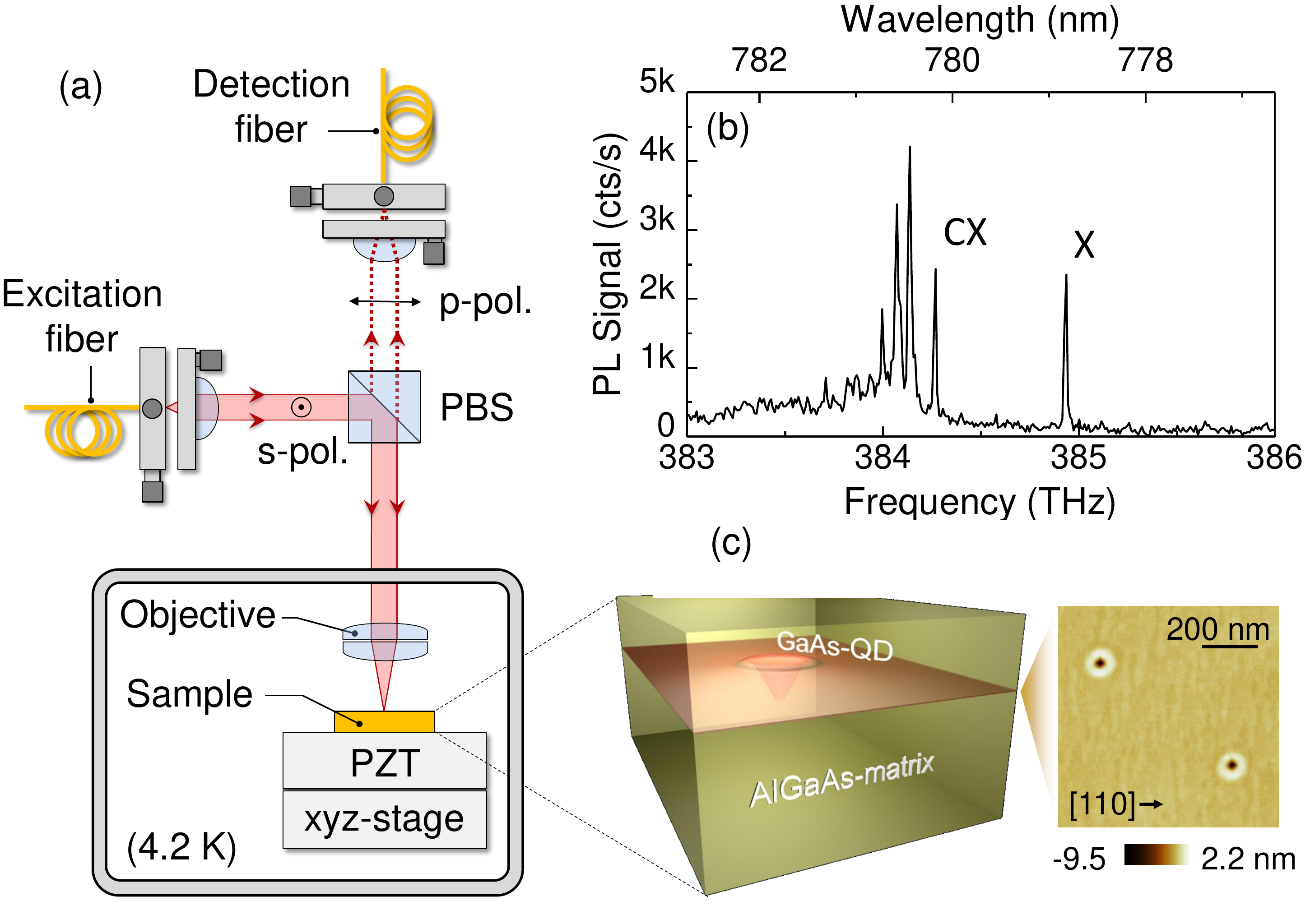}
\caption{The experimental setup. a) Schematics of the resonance fluorescence setup showing orthogonally polarized excitation and detection. PBS refers to a polarizing beam splitter (PBS). The sample is glued to a piezo-electric transducer (PZT) and mounted onto an xyz-positioning stage. A solid immersion lens (SIL) on the surface of the sample increases the collection efficiency. b) PL spectrum of a single QD under non-resonant excitation at 633 nm ($I_{\text{NR}} \sim 7$ \textmu W/\textmu m$^2$). We identify the neutral exciton (X) and a charged exciton (CX) which display narrow linewidths, limited here by the $9$ GHz spectrometer resolution. c) Sketch of the QD layer and an AFM picture of the nano-holes obtained with in-situ etching \cite{Rastelli2004}.}
\label{Fig1}
\end{figure}

Our solid-state source of single photons, Fig.\ \ref{Fig1}, consists of a GaAs/AlGaAs quantum dot (QD) obtained by filling Al-droplet-etched nanoholes with GaAs \cite{Huo2013}.
The holes are formed by depositing 0.5 mono-layer (ML) of aluminium at a growth rate of 0.5 ML/s and at a temperature of 600~\textdegree C on a Al$_{0.4}$Ga$_{0.6}$As surface. This is followed by a 5 minute annealing step in arsenic ambiance. The holes are then filled with GaAs grown at 0.1 ML/s and capped again with Al$_{0.4}$Ga$_{0.6}$As resulting in strain-free GaAs QDs.
The photoluminescence (PL) from the ensemble is adjusted to $\sim 780$ nm, the wavelength of the Rb D2-line, by controlling the exact amount of deposited GaAs. Fig.\ \ref{Fig1}b shows a typical PL spectrum from a single QD recorded at $4.2$~K with non-resonant excitation at 633 nm. We observe several lines in the PL spectrum. We identify in particular the neutral exciton (X) and a red-detuned charged exciton (CX). The other lines are related to other exciton states, as yet unidentified. To fine tune the QD frequency with respect to the Rb transition lines, the sample is glued onto a piezo-electric transducer which induces uniaxial strain in the sample \cite{Seidl2006a,Kumar2011}, Fig.\ \ref{Fig1}a. By scanning the piezo-voltage, reversible tuning over $30$ GHz is achieved with very little creep from the piezo-electric elements, see Fig.\ \ref{Fig2}c. In fact, the emission frequencies of the PL lines are stable over the course of a day such that a stabilization scheme was not necessary in these experiments.

\section{Resonance fluorescence on a single QD}
We first report resonance fluorescence on a single GaAs QD, the artificial Rb atom. For this, we use the dark-field microscope sketched in Fig.\ \ref{Fig1}a. A resonant laser beam is focussed onto the sample with linear polarization; resonance fluorescence from the QD is detected in the orthogonal polarization \cite{Kuhlmann2013b}. Careful control of the polarization suppresses the back-scattered laser light by 80 dB. We find that very weak non-resonant laser light ($\lambda = 633$ nm, $I_{\rm NR} \gtrsim 0.8$ nW/$\mu$m$^2$) is a necessary condition to observe resonance fluorescence on CX. This non-resonant excitation quenches the excitation of the neutral X and therefore acts as an ``optical gate" \cite{Nguyen2012}. This result was reproducibly observed on all five QDs that we tested. 

\begin{figure}[tb]
\includegraphics[width=85mm]{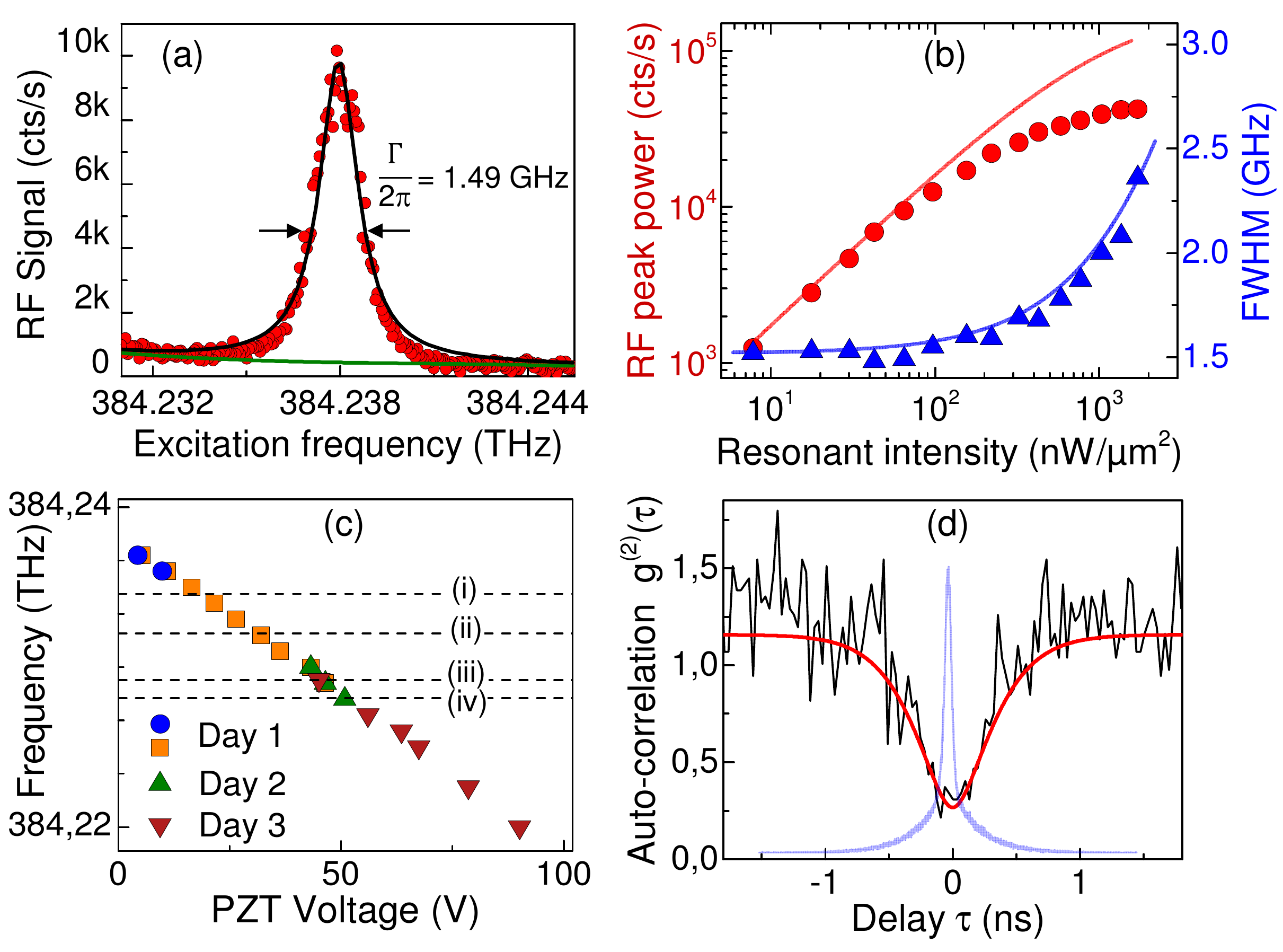}
\caption{Resonance fluorescence of the charged exciton, QD1. a) Resonance fluorescence spectrum in the low power regime ($I_{\text{R}}=16$ nW/$\mu$m$^{2}$). The laser background ($\leq 780$ cts/s over the $12$ GHz scanning range) is indicated in green. b) Resonance fluorescence intensity and FWHM as a function of resonant laser power. c) Frequency tuning of CX showing a linear response to the applied voltage with very little creep over the course of several days. The D2 transitions of Rb are indicated as dashed lines: (i) $^{87}$Rb $F_g =1 \rightarrow F^{'}_e$, (ii) $^{85}$Rb $F_g =2 \rightarrow F^{'}_e$, (iii) $^{85}$Rb $F_g =3 \rightarrow F^{'}_e$ and (iv) $^{87}$Rb $F_g =2 \rightarrow F^{'}_e$. d) Second order correlation of the resonance fluorescence signal. In blue, the detectors' response function (arbitrary units for the y-axis) measured with ultra-short laser pulses ($5$ ps) at the QD frequency. The red line results from a fit using Eq.~\ref{g2TLS} convoluted with the detectors' response function. All data are obtained in the presence of an additional weak, constant non-resonant laser excitation of $I_{\text{NR}} \approx 0.8$ nW/\textmu m$^{2}$. The background associated with the non-resonant excitation is smaller than the detectors' dark counts.}
\label{Fig2}
\end{figure} 

To record resonance fluorescence spectra, we monitor the count rate on a CCD camera as  we sweep the laser frequency across the QD transition, as illustrated in Fig.\ \ref{Fig2}a for the CX transition of QD1. The spectrum is fitted with a Lorentzian profile, and displays a  signal-to-background ratio $\text{S:B} > 23$ at the resonance. In the low power regime, the linewidth is $\Gamma / 2\pi= 1.49 \pm 0.04$ GHz, see Fig.\ \ref{Fig2}a. We confirm the anti-bunched nature of the emitted photons by performing second-order correlation measurements on the resonance fluorescence signal, Fig.\ \ref{Fig2}d. There is a small bunching on the normalized data  ($g_2(\tau) = 1.25$ for $\tau > 1$ ns) which results from a slow blinking process, discussed below. For $\tau \ll 1\,\mu$s, the exact blinking dynamics can be ignored and the data are fitted to the product of a constant pre-factor, which accounts for the QD dead-time (i.e.\ the blinking), and the second-order correlation function of a resonantly driven two-level system \cite{Flagg2009}
\begin{eqnarray}
\label{g2TLS}
g^{(2)}_{\text{TLS}}(\tau) & = &  1- e^{-\frac{1}{4}(3\Gamma_{\text{sp}} +2 \gamma^{*})\tau} \nonumber \\ 
& & \times \left(\cos{\lambda \tau} +\frac{3 \Gamma_{\text{sp}} + 2\gamma^{*}}{4 \lambda} \sin{\lambda \tau}\right)  
\end{eqnarray}
where $\Gamma_{\text{sp}}$ is the spontaneous radiative emission rate, $\gamma^{*}$ corresponds to the pure dephasing rate and $\lambda = \sqrt{\Omega^{2} - \frac{1}{16} \left( \Gamma_{\text{sp}} - 2 \gamma^{*} \right)^{2}}$, with $\Omega$ the Rabi frequency of the resonant drive. Taking the experimentally measured response of the detectors into account, we find a very nice agreement and thus a coincidence detection probability consistent with zero at zero delay, the signature of pure single photon emission. \\


\section{Spectroscopy of the Rubidium atomic ensemble with QD photons}

\begin{figure}[]
\includegraphics[width=85mm]{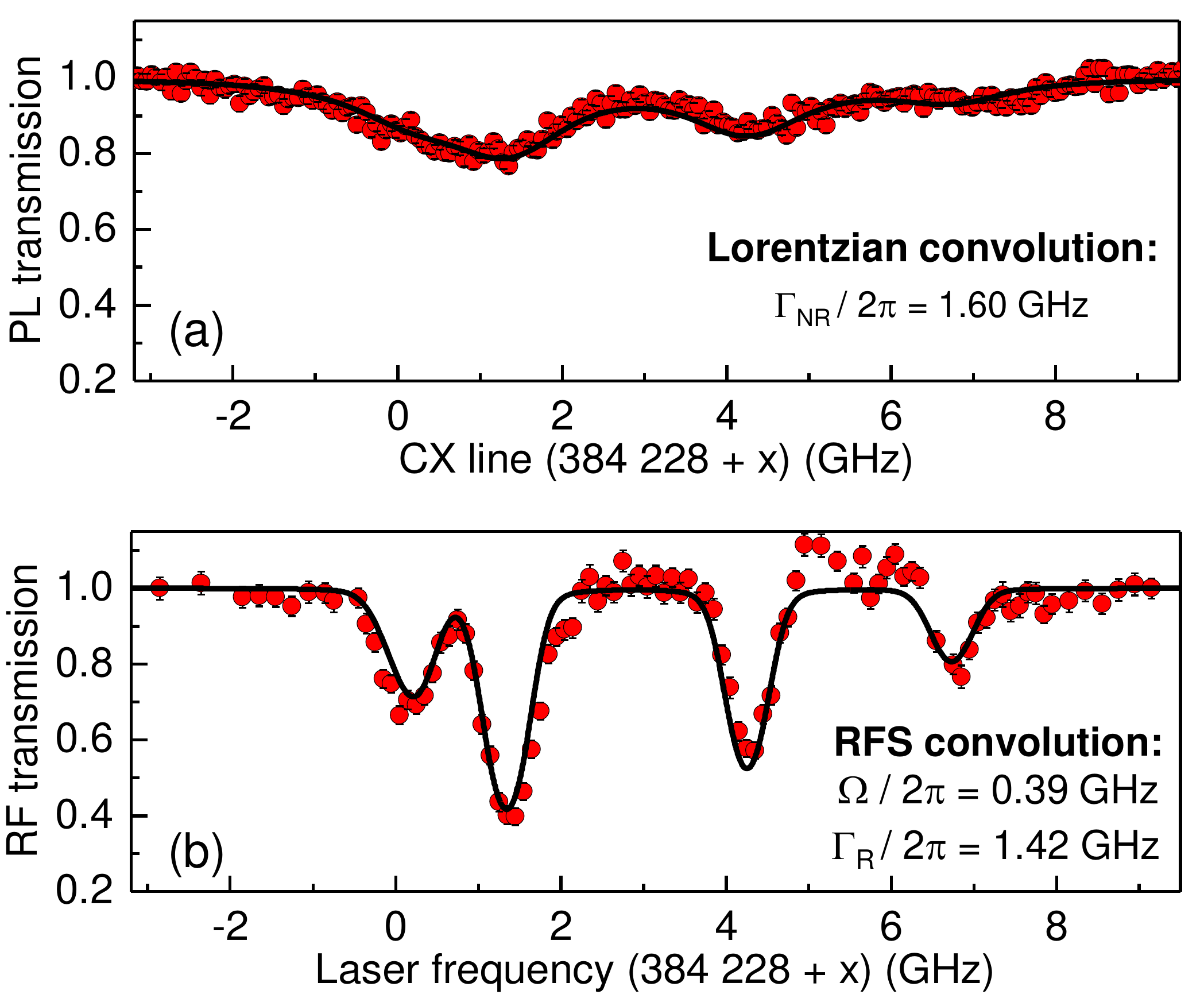}
\caption{Spectroscopy of the Rb D2-transitions using QD photons. In a) QD1 is excited non-resonantly and the CX resonance is swept through the Rb transitions. The solid line is a fit based on the convolution between the atomic transmission spectrum and a Lorentzian line accounting for the spectral width of the QD photons. In b) CX is driven at resonance in the coherent Rayleigh scattering regime. The solid line is a fit where the QD is modeled as a two-level scatterer with associated resonance fluorescence spectrum (RFS).}
\label{Fig3}
\end{figure}

We now turn to the spectroscopy of the Rb atomic ensemble using QD photons. We insert a room temperature $75$ mm long Rb vapor cell in the detection line. The cell contains both $^{85}$Rb and $^{87}$Rb in natural abundance ($72.2 \%$ and $27.8 \%$, respectively). In a first experiment, QD1 is excited with the non-resonant pump only with $I_{\text{NR}} = 7.1$ \textmu W/\textmu m$^2$, Fig.\ \ref{Fig3}a. Transmission through the atomic cloud is recorded as the piezo-voltage is increased thus tuning the QD emission frequency. As the CX transition is scanned from $384.225$ THz to $384.237$ THz, we observe several dips in the transmission corresponding to the hyperfine structure of the two rubidium isotopes, Fig.\ \ref{Fig2}c. In order to distinguish between the QD and the atomic contributions to the linewidth, we perform a calibration measurement on the vapor cell by measuring the transmission with the laser only (FWHM $\leq 1$ MHz @100$\mu$s). The result, shown in the appendix (see Fig.\ \ref{Fig:VaporAbsorptionSpectrum}), is fitted to the theoretical Rb transmission spectrum, where the only unknown is the vapor cell temperature. Excellent agreement is found for $T= 24.8$ \textdegree C, corresponding to a Doppler broadening of $510$ MHz. To describe the transmission spectrum recorded with QD photons, we then convolve the Rb spectrum with a Lorentzian profile of width $\Gamma_{\text{NR}}$, the QD linewidth under non-resonant excitation. Best agreement between the resulting function and the data is obtained for $\Gamma_{\text{NR}}/ 2\pi = 1.60 \pm 0.20$ GHz. The modest depth of the transmission peaks reflects the mismatch between the QD linewidth and the atomic spectral width.  

A lifetime-limited linewidth implies a negligible rate of exciton dephasing in the QD. In turn, this opens the possibility of generating single photons by coherent Rayleigh scattering. The resonance fluorescence can be divided into a coherent part, the Rayleigh scattering of the incoming laser light, and an incoherent part, resulting from an absorption and re-emission cycle. Including pure dephasing, the fraction of coherently scattered photons is given by
\begin{equation}
\frac{I_{\text{coherent}}}{I_{\text{total}}}=\frac{\Gamma_{\text{sp}}^{2}}{2\Omega^{2}+\Gamma_{\text{sp}}^{2}+ 2\gamma^{*}\Gamma_{\text{sp}}}.
\end{equation}
(See Appendix \ref{Apdx_theory_QDres} for a complete description of the resonant spectrum.) The ratio is maximum in the low power regime ($\Omega \ll \Gamma_{\text{sp}}$), the Rayleigh regime, and approaches unity should $\gamma^*$ become negligible compared to $\Gamma_{\text{sp}}$. The last point highlights the importance of achieving a small dephasing rate. Conversely, the ratio decreases at high power where the strong excitation leads to inelastic scattering (Mollow triplet). We explore the possibility of coherent Rayleigh scattering in a second experiment where we drive the QD resonantly in the low power limit ($I_{\text{R}} = 141$ nW/$\mu$m$^2$). The resulting Rb transmission spectrum is shown in Fig.\ \ref{Fig3}b. For a given driving laser frequency, we tune the QD into resonance via the piezo-voltage, and we measure the resonance fluorescence signal transmitted through the Rb vapor cell. This is then repeated for different laser frequencies. The transmission data are normalized using a linear baseline defined by points recorded when the QD is detuned from the Rb transitions. In Fig.\ \ref{Fig3}b, the four dips corresponding to the D2-transitions of $^{85}$Rb and $^{87}$Rb can now clearly be resolved, showing negligible broadening of the atomic transitions beyond that of the atomic vapor itself. This implies that the spectrum of the light scattered by the QD has been narrowed down significantly below the lifetime limit, a clear evidence of coherent scattering from the QD \cite{Matthiesen2012}.

To fit the measured spectrum in the Rayleigh regime, we compute the convolution between the atomic spectrum and the resonant emission spectrum, with $\Gamma_{\text{sp}}$, $\gamma^*$ and $\Omega$ as free parameters. In order to determine a value for each parameter with the highest accuracy, we perform a global fit on both the transmission spectrum (Fig.\ \ref{Fig3}b) and the second-order correlation measurement (Fig.\ \ref{Fig2}d). From this combined analysis we determine $\Gamma_{\text{sp}} / 2\pi= 1.42 \pm 0.12$ GHz, $\gamma^*/ 2\pi = 0 \pm \binom{0.10}{0}$ GHz and $\Omega / 2\pi = 0.39 \pm 0.10$ GHz, which corresponds to a fraction of coherently scattered photons as high as $87 \%$ (see details in Appendix \ref{Apdx_theory_QDres}). These results are further supported by recording a decay curve following non-resonant pulsed excitation. The data, which, incidentally, point to an unusually slow relaxation mechanism for transferring carriers from high energy continuum states into the QD, result in $\Gamma_{\text{sp}} / 2\pi= 1.7 \pm 0.2$ GHz, consistent with the spectroscopy analysis (see Fig.\ \ref{Fig:Lifetime_pulsed_NR} from Appendix~\ref{Apdx_NRdecay}). We note also the excellent agreement with the power broadening experiment where the resonance fluorescence linewidth is described within the two-level system framework, with $\Gamma_{\text{sp}}$ and $\gamma^*$ as input parameters, Fig.\ \ref{Fig2}b. 

These results allow us to make an important conclusion, namely that we achieve lifetime-limited emission with our artificial atom. We thus combine, in a solid-state environment, a high single photon flux with negligible dephasing, a key result for further quantum optics experiments, for instance the generation of indistinguishable photons. In addition, this conclusion applies not only under resonant excitation (resonance fluorescence), but also under non-resonant excitation (photoluminescence). This is a surprising result in the context of InGaAs QDs where the transform limit has been achieved only with resonant excitation and for very specific conditions \cite{Kuhlmann2015}; in the best case with non-resonant excitation the linewidth is about a factor of two larger than the transform limit \cite{Bayer2002} and is typically much larger still. These exceptional results on GaAs QDs reflect the high quality of the epitaxial material combined with the short radiative lifetime and possibly an unknown semiconductor advantage of strain-free QDs over highly-strained QDs.\\

\section{Blinking in the QD signal}
\begin{figure}[tb]
\includegraphics[width=85mm]{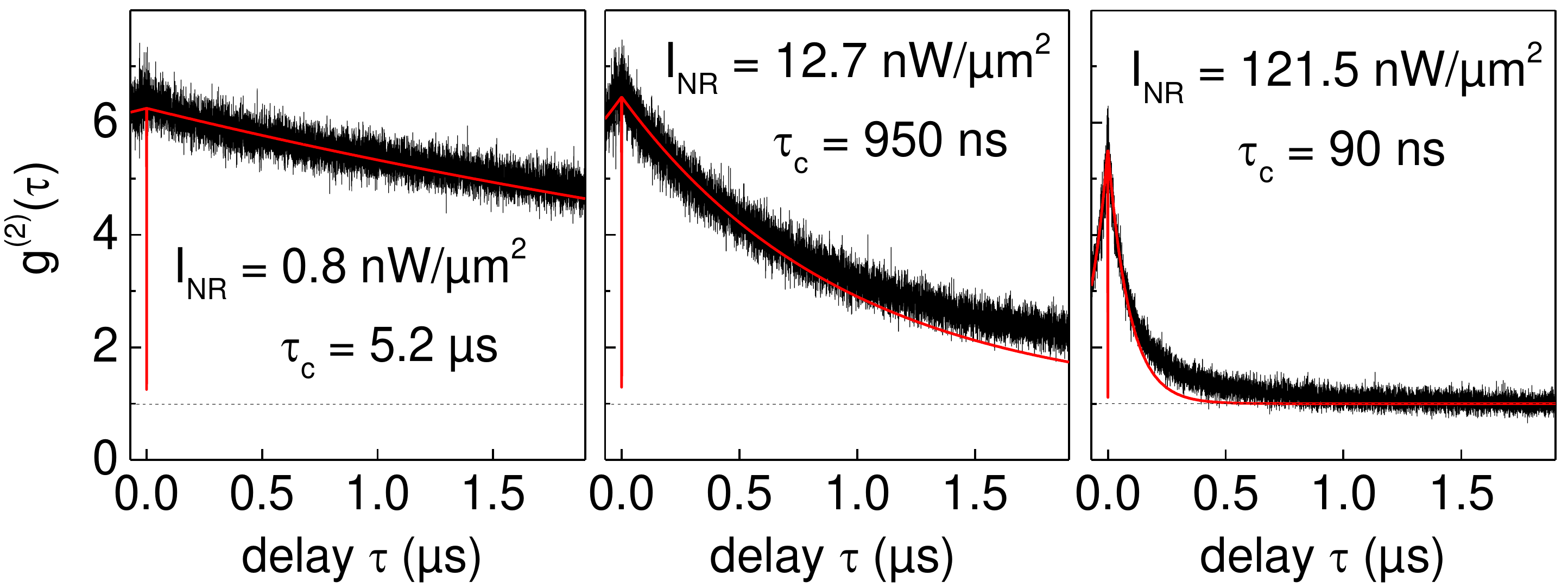}
\caption{Blinking statistics on QD2. Second order correlation measurement of the resonance fluorescence signal of CX as a function of increased non-resonant pump ($I_{\text{R}} = 180.5$ nW/\textmu m$^2$). Solid red lines are fits obtained from Eq. (\ref{Eq:g2blink}).}
\label{Fig4}
\end{figure}

The solid-state environment results in negligible dephasing of the QD single photon source. However, the effects of the solid-state environment are not completely suppressed: Fig.\ \ref{Fig4} shows a correlation measurement under resonant excitation on a second QD for three different values of non-resonant power. The data are normalized to the average count per time bin for a Poissonian source, $N = N_1 N_2 \tau_{b} T$, with $N_1$ and $N_2$ the count rates on each avalanche photo-diode, $\tau_b$ the time-resolution of the experiment and $T$ the total integration time. In addition to the anti-bunching at zero delay already outlined in Fig.\ \ref{Fig2}c, we observe a strong bunching peak at short delays ($g^{2}(\tau)$ as high as $6.5$). This corresponds to the signature of blinking in the QD emission \cite{Michler2000b}: the presence of dead times in the QD fluorescence produces packets of single photons separated in time. Assuming a simple Boolean statistics for the blinking process \cite{Machlup1954}, ergodic and statistically independent of the two-level radiative decay, the second-order correlation function of the QD signal can be expressed as
\begin{equation}
\label{Eq:g2blink}
g^{(2)}(\tau) = \left(1 + \frac{1-\beta}{\beta} e^{- \tau / \tau_c } \right) \times g^{(2)}_{\text{TLS}}(\tau)
\end{equation}
where $\beta$ corresponds to the fraction of time in which the QD is in an ``on" state, and $\tau_c$ to the correlation time of the blinking process. The first term (left bracket) accounts for telegraph noise associated with the blinking, the second term for the dynamics of the resonantly driven two-level system, cf. Eq.\ (\ref{g2TLS}). From the fit of the data, we extract $\beta_{CX} \sim 16 \%$, a less favorable situation for the charged exciton in QD2 as compared to QD1 ($\beta_{CX} \sim 80\%$, see Fig.\ \ref{Fig2}d and Fig.\ \ref{g2_QD1_longtimes} from Appendix \ref{Apdx_g2QD1long}). The blinking dynamics are strongly modified as we increase the non-resonant power. We find that $\tau_c$ varies by several orders of magnitude over the available range of power with $\beta$ remaining approximately constant. This result was reproducibly observed on all QDs we tested and reflects the general nature of the solid-state environment. It shows how the non-resonant laser power offers some control over the environment, here in all likelihood fluctuations in charge (either in the QD or in the immediate vicinity of the QD) which bring the QD in and out of resonance in a telegraph fashion with the fixed frequency laser. We note that $\tau_c$ is in all cases considerably larger than the radiative lifetime (90~ps) such that the blinking contribution to the QD linewidth is small: the telegraph noise is consistent with the claim of a lifetime-limited QD linewidth. Also, we note that the simple on:off model does not capture all the details of the blinking dynamics. At high resonant power, the decrease in resonance fluorescence peak signal at the highest resonant powers (Fig.\ \ref{Fig2}b) is probably related to an increase in the QD dead-time.

\section{Conclusion}
In conclusion, we report here a quantum hybrid system consisting of a frequency-matched solid-state source of single photons, a single quantum dot, and a Rb atomic vapor. The quantum dots exhibit lifetime-limited linewidths, even under non-resonant excitation. Resonance fluorescence in the Rayleigh scattering regime is used to address the bandwidth mismatch between the two quantum systems. The most significant solid-state noise is at $\sim$ MHz frequencies and results in telegraph noise in the emission reflecting QD blinking. We demonstrate some control over this correlation time, useful in the context of decoupling the QD from its complex environment. Further work should address this noise and also engineering of the photonic environment in order to achieve a higher QD single photon collection efficiency. Implementation of quantum memory protocols can then be attempted \cite{Rakher2013}.

\begin{acknowledgments}
We acknowledge financial support from NCCR QSIT and  the European Union Seventh Framework Programme 209 (FP7/2007-2013) under Grant Agreement No. 601126 210 (HANAS), and thank M. Rakher and N. Sangouard for fruitful discussions. J-P.J., M.M. and L.B. made equal contributions to this work. 
\end{acknowledgments}

\appendix

\section{ Vapor cell absorption spectrum}

\subsection{Theory}

\begin{figure*}[]
  \begin{minipage}[b]{0.2\linewidth}
    \centering
   {\footnotesize (a)}\\
    \vspace{8pt}
   	\includegraphics[width=0.67\textwidth]{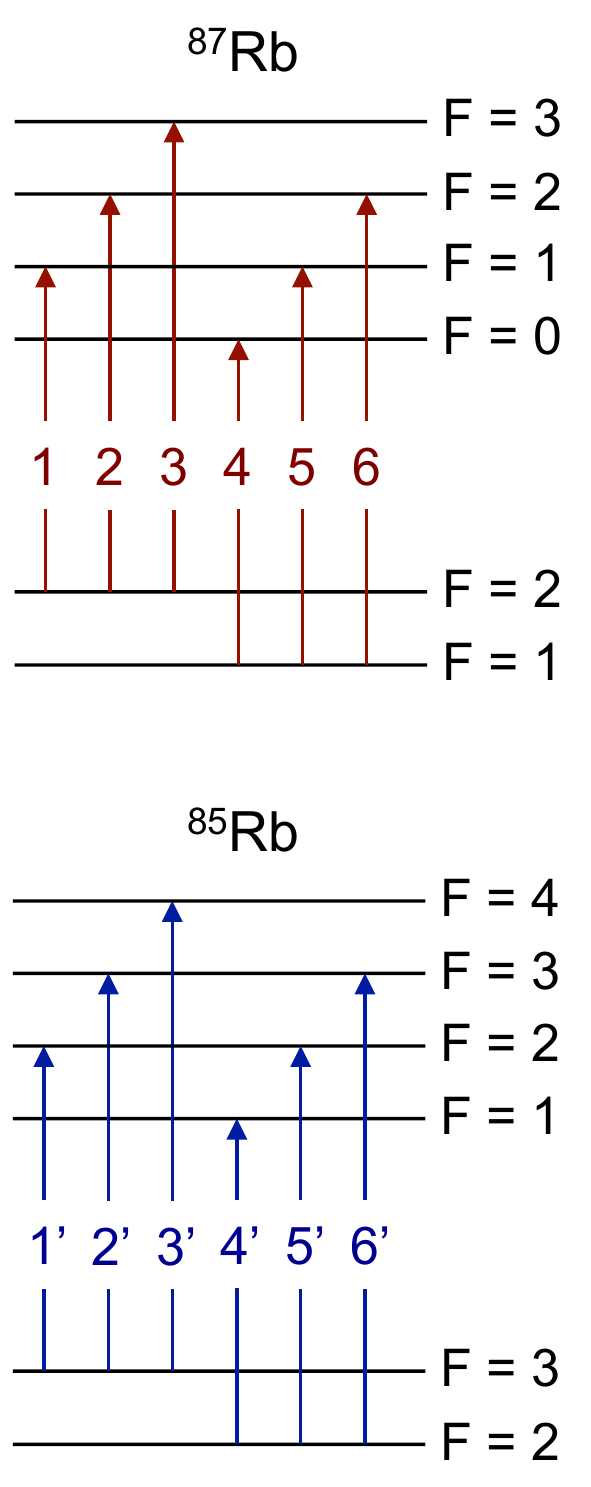}
    \par\vspace{0pt}
  \end{minipage}
  \begin{minipage}[b]{0.79\linewidth}
    \centering
    {\footnotesize (b)}\\
    \vspace{14pt}
    \begin{adjustbox}{max width=0.99\textwidth}
    \begin{tabular}{  c | c | c | c  }   
          Isotope                        ~   &   ~~ Transition $j$ ~~  & ~~  $(\omega_j-\omega_{\rm ref})/2\pi$ (MHz) ~~  & ~~Strength factor $C_j^2$ ~~ \\ \hline
        \multirow{6}{*}{${\rm ^{87}Rb}$} ~   &   ~~    1    ~~      & ~~    -156.941   ~~   &  $1/18$              \\ \cline{2-4}     
                                         ~   &   ~~    2    ~~      & ~~       0       ~~   &  $5/18$              \\ \cline{2-4} 
                                         ~   &   ~~    3    ~~      & ~~     266.652   ~~   &  $7/9$               \\ \cline{2-4}        
                                         ~   &   ~~    4    ~~      & ~~    6605.520   ~~   &  $1/9$               \\ \cline{2-4}
                                         ~   &   ~~    5    ~~      & ~~    6677.742   ~~   &  $5/18$              \\ \cline{2-4}
                                         ~   &   ~~    6    ~~      & ~~    6834.683   ~~   &  $5/18$              \\ \hline    
		\hline    
        \multirow{6}{*}{${\rm ^{85}Rb}$} ~   &   ~~    1    ~~      & ~~    1207.094   ~~   &  $10/81$             \\ \cline{2-4}
                                         ~   &   ~~    2    ~~      & ~~    1270.494   ~~   &  $35/81$             \\ \cline{2-4} 
                                         ~   &   ~~    3    ~~      & ~~    1391.134   ~~   &    $1$               \\ \cline{2-4}         
                                         ~   &   ~~    4    ~~      & ~~    4213.453   ~~   &   $1/3$              \\ \cline{2-4} 
                                         ~   &   ~~    5    ~~      & ~~    4242.826   ~~   &  $35/81$             \\ \cline{2-4}                                                     
                                         ~   &   ~~    6    ~~      & ~~    4306.226   ~~   &  $28/81$             \\      
    \end{tabular}
    \end{adjustbox}
    \par\vspace{0pt}

    \end{minipage}
    \caption{{\bf Hyperfine structure of \boldmath${\rm ^{87}Rb}$ and \boldmath${\rm ^{85}Rb}$ D2 line.} (a) Sketch of the allowed hyperfine transitions. (b) Properties of the hyperfine transitions. Frequencies are given with respect to the ${\rm ^{87}Rb}$ transition $F_g=2\rightarrow F_e=2$ of angular frequency $\omega_{\rm ref}=2\pi \times$~384~227~848.551~MHz. Transition strength factors $C_j^2$ are computed for linearly polarized incident light.}
    \label{Fig:D2Properties}
  \end{figure*}

We derive here the absorption spectrum of the Rb vapor cell, following the method described in Ref.~\onlinecite{Siddons2008}. For weak probe intensity, the transmission of a monochromatic wave of angular frequency $\omega$ through an atomic vapor with uniform density is given by
\begin{equation}
\label{Eq:VaporTransmission}
\mathcal{T}_{\rm vapor}(\omega,T)= e^{-\alpha(\omega,T) L},
\end{equation}
where $L$ is the length of the vapor cell and $\alpha(\omega,T)$ is the absorption coefficient of the atomic vapor, which is only dependent on the temperature $T$. Our cell contains ${\rm ^{85}Rb}$ and ${\rm ^{87}Rb}$ in natural abundance ($\epsilon_{85}=72.17$\,\% and  $\epsilon_{87}=27.83$\,\%) so that the total absorption reads $\alpha(\omega,T) = \alpha_{\rm 85}(\omega,T) + \alpha_{\rm 87}(\omega,T)$. For each isotope, we consider the six allowed electric dipole hyperfine transitions shown in Fig.~\ref{Fig:D2Properties}a, which leads to the following expression for the absorption of isotope $i$
\begin{equation}
\label{Eq:AbsorptionCoefficient}
\alpha_{\rm i}(\omega,T) = \sum_{j=1}^{6} \frac{n_{\rm i}(T)}{2(2 \, \mathcal{I}_i+1)~\hbar \epsilon_0} C_j^2 d^2 \times s^i_{\Gamma}(\omega-\omega_j,T),
\end{equation}
where $d=5.177~e\,a_0$ (with $a_0$ the Bohr radius) is the reduced dipole matrix element computed for the D2 line, $C_j^2=\sum_{m_F} c_j^2$ is the total strength coefficient of the degenerate hyperfine transition $j$ (tabulated in Fig.~\ref{Fig:D2Properties}b for linear incident polarization) and $n_{\rm i}(T)/[2(2\,\mathcal{I}_i+1)]$ is the isotope atomic density per Zeeman sublevel. ${\rm ^{85}Rb}$ and ${\rm ^{87}Rb}$ have nuclear spins $\mathcal{I}_{\rm 85}=5/2$ and $\mathcal{I}_{\rm 87}=3/2$ and their relative density $n_{\rm i}(T)=\epsilon_i \, n(T)$ is obtained from the ideal gas law where the vapor pressure $p(T)$ is given by equations (A.1) and (A.2) of Ref.~\cite{Siddons2008}. Finally, the lineshape factor 
\begin{multline}
\label{Eq:VoigtProfileNorm}
s^i_{\Gamma}(\delta_j,T) = \int_{-\infty}^{+\infty} \frac{\Gamma/2}{(\Gamma/2)^2+(\delta_j-k\,v)^2}\\
\times \frac{1}{\sqrt{\pi} \sigma_{i}(T) }  \exp\left(-\frac{v^2}{\sigma^2_{i}(T)}\right) \mathrm{d}v,
\end{multline}
corresponds to the Doppler broadened profile of the atomic transition $j$. We take the Lorentzian profile of the atom with natural linewidth $\Gamma = 2 \pi \times 6.065$~MHz (the experimentally measured decay rate of the $5^{2}P_{3/2}$ atomic state~\cite{Steck}) integrated over the Gaussian distribution of atomic velocities parallel to the probe beam, with $1/e$ width $\sigma_{i}(T)=\sqrt{2 k_B T/m_i}$ ($k_B$ is the Boltzmann constant, $m_i$ is the isotope atomic mass). At $T = 24.8$\,\textdegree C, the thermal longitudinal motion of the atom leads to a full-width half maximum (FWHM) Doppler broadening $\Delta \omega~=~2\sqrt{\ln 2} \, \omega \, \sigma_{i}/c \simeq 2 \pi \times 0.51$ GHz for the D2 line at 780\,nm.

\subsection{Experiment}
\label{Apdx_Experiment}
\begin{figure}[h]
\includegraphics[width=85mm]{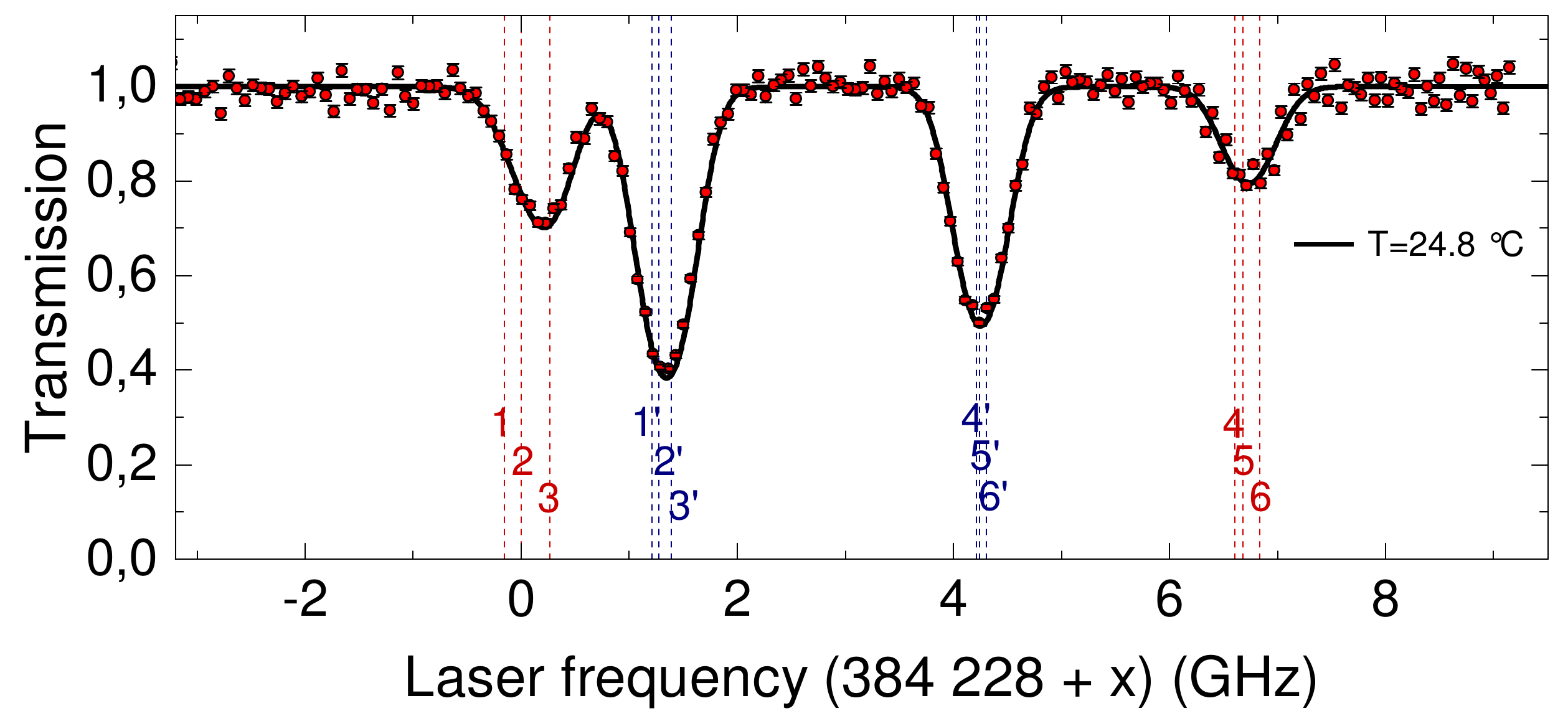}
\caption{{\bf Transmission of the Rb vapor cell.} The laser intensity is adjusted to the typical QD resonance fluorescence level of 5\,kcts/s. The exposure time is 1\,s per data point. Raw measurements are normalized using a linear baseline. The solid black line is a fit using equations~(\ref{Eq:VaporTransmission}), (\ref{Eq:AbsorptionCoefficient}) and (\ref{Eq:VoigtProfileNorm}) with $T=24.8\pm 0.2$\,\textdegree C.}
\label{Fig:VaporAbsorptionSpectrum}
\end{figure}

 Fig.~\ref{Fig:VaporAbsorptionSpectrum} shows an experimental transmission spectrum of a 75\,mm rubidium vapor cell measured using a tunable 780\,nm external cavity diode laser (short term [$100\,\mu$s] FWHM $< 1$ MHz) with linear incident polarization. The data are fitted using equations~(\ref{Eq:VaporTransmission}), (\ref{Eq:AbsorptionCoefficient}) and (\ref{Eq:VoigtProfileNorm}), where the vapor temperature is the only free parameter. Excellent agreement is obtained for $T=24.8 \pm 0.2$\,\textdegree C (see solid line).

\section{Theory of the QD response to a resonant field}
\label{Apdx_theory_QDres}

\subsection{First order coherence $g^{(1)}(\tau)$ and power spectrum $S(\omega_{\rm sc})$} 

We aim at describing the resonance fluorescence (RF) power spectrum $S(\omega_{\rm sc})$ of a QD excited resonantly. To do so, we assume that the QD behaves as a two-level system. We follow the approach of Mollow\cite{Mollow1969} and extend it to include the additional pure dephasing associated to the extra coupling to the QD solid-state environment. We first evaluate the first order coherence $g^{(1)}(t,\tau)$ of the field scattered by the QD, from which we can easily derive its power spectrum. 

The two-level system has a ground state $\ket{g}$, excited state $\ket{e}$ (decay rate $\Gamma_{\rm sp}$), and a transition angular frequency $\omega_0$. Neglecting retardation effects, the first-order coherence reads
\begin{equation}
\label{Eq:FirstOrderCoherence1} 
g^{(1)}(t,\tau) = \frac{\langle \hat{\pi}^{\dag}(t) \hat{\pi}(t+\tau) \rangle }{\langle \hat{\pi}^{\dag}(t) \hat{\pi}(t) \rangle},
\end{equation}
\vspace{-1.2cm}
\begin{equation}
{\rm with~~~~}
\left \lbrace
\begin{array}{c c c c l}
\hat{\pi}^{\dag} &=& \ket{e}\bra{g},~~~~ \langle \hat{\pi}^{\dag}(t) \rangle &=& \tilde{\rho}_{ge}(t) e^{i \omega t}\\
\hat{\pi}        &=& \ket{g}\bra{e},~~~~ \langle \hat{\pi}(t) \rangle &=& \tilde{\rho}_{eg}(t) e^{-i\omega t},
\end{array}
\right.
\end{equation}
where $\rho_{ij}$  are the density matrix elements of the two-level system, and $\pi^{\dag}$ and $\pi$ are atomic transition operators. The dynamics under coherent illumination are described by the optical Bloch equations\cite{Loudon}. The steady-state expectation values of the transition operators are computed in the interaction picture using the quantum regression theorem. The decay rates are $1/T_1 = \Gamma_{\rm sp}$ for the populations, and $1/T_2 = \Gamma_{\rm sp}/2 + \gamma^*$ for the coherences. Following the derivation of Ref.~\onlinecite{Mollow1969} we obtain the steady-state expression $g^{(1)}(\tau)=\lim\limits_{t \to \infty} g^{(1)}(t,\tau)$, which, in the resonant case $\omega=\omega_0$, is given by
\begin{multline}
g^{(1)}(\tau) \, e^{i \omega \tau}  = \frac{\Gamma_{\rm sp}^{2}}{2\Omega^{2}+\Gamma_{\rm sp}^{2}+2\gamma^{*}\Gamma_{\rm sp}} +\frac{1}{2} e^{-(\frac{\Gamma_{\rm sp}}{2}+\gamma^*)\tau}\\
+  e^{-(\frac{3 \Gamma_{\rm sp} + 2\gamma^*}{4})\tau} \left[ \frac{P}{2} \cos \lambda \tau - \frac{Q}{2} \sin \lambda \tau \right], 
\label{Eq:FirstOrderCoherence2}
\end{multline}
with
\begin{equation*}
\begin{array}{l l}
 \left \lbrace 
\begin{array}{c l l}
\lambda &=& \sqrt{\Omega^{2} - \left(\frac{\Gamma_{\rm sp}}{4}-\frac{\gamma^{*}}{2}\right)^{2}} \\[\bigskipamount]
P       &=& \dfrac{2\Omega^2-\Gamma_{\rm sp}+2\gamma^*\Gamma_{\rm sp}}{2\Omega^2+\Gamma_{\rm sp}+2\gamma^*\Gamma_{\rm sp}} \\[\bigskipamount]
Q       &=& \dfrac{\Omega^2(5\Gamma_{\rm sp}-2\gamma^*) - 2\gamma^{*2}\Gamma_{\rm sp} + 2\gamma^*\Gamma_{\rm sp}^2-\Gamma_{\rm sp}^3/2}{2\lambda (2\Omega^2+\Gamma_{\rm sp}+2\gamma^*\Gamma_{\rm sp})}.
\end{array}
\right.
\end{array}
\label{Eq:FirstOrderCoherence3}
\end{equation*}
The Fourier transform of $g^{(1)}(\tau)$ gives the expression for the RF power spectrum
\begin{multline}
\label{Eq:RFspectrum}
S(\omega_{\rm sc}) = \frac{\Gamma_{\rm sp}^{2}}{2\Omega^{2}+\Gamma_{\rm sp}^{2}+ 2\gamma^{*}\Gamma_{\rm sp}} \delta(\omega_{\rm sc}-\omega_0)\\
+ \frac{1}{2\pi}\frac{\frac{\Gamma_{\rm sp}}{2}+\gamma^{*}}{(\omega_{\rm sc}-\omega_0)^{2} + (\frac{\Gamma_{\rm sp}}{2}+\gamma^{*)^2}} \\
+\frac{1}{4\pi}\frac{\left(\frac{3}{4}\Gamma_{\rm sp}+\frac{1}{2}\gamma^*\right)P - (\omega_{\rm sc} -\omega_0 -\lambda) Q}{(\omega_{\rm sc} -\omega_0 - \lambda)^2+\left(\frac{3}{4}\Gamma_{\rm sp}+\frac{1}{2}\gamma^*\right)^2}\\
+\frac{1}{4\pi}\frac{\left(\frac{3}{4}\Gamma_{\rm sp}+\frac{1}{2}\gamma^*\right)P + (\omega_{\rm sc} - \omega_0 +\lambda) Q}{(\omega_{\rm sc} -\omega_0 + \lambda)^2+\left(\frac{3}{4}\Gamma_{\rm sp}+\frac{1}{2}\gamma^*\right)^2},
\end{multline}
which depends on three parameters only: the driving Rabi frequency $\Omega$, the radiative decay rate of the excited state $\Gamma_{\rm sp}$, and the pure dephasing rate $\gamma^{*}$. Experimental RF spectra result from the convolution of (\ref{Eq:RFspectrum}) with the emission spectrum of the resonant laser. In practice, we use a highly coherent 780\,nm external cavity diode laser, that we model by a Gaussian profile with a full width at half maximum of 1\,MHz.

\subsection{Second order coherence $g_{\rm TLS}^{(2)}(\tau)$} 

Within the two-level system model (TLS), the second order coherence of the field scattered by the QD is given by
\begin{equation}
\label{Eq:SecondOrderCoherence1}
g^{(2)}_{\rm TLS}(t,\tau) = \frac{\langle \hat{\pi}^{\dag}(t) \hat{\pi}^{\dag}(t+\tau) \hat{\pi}(t+\tau) \hat{\pi}(t) \rangle }{\langle \hat{\pi}^{\dag}(t) \hat{\pi}(t) \rangle^2}.
\end{equation}
As before, it is derived in the interaction picture using the quantum regression theorem. Using the same notations, we find
\begin{equation}
\label{Eq:g2TLSb}
g^{(2)}_{\rm TLS}(\tau) =  1- e^{-\frac{3\Gamma_{\rm sp} +2 \gamma^{*}}{4}\tau} \left(\cos{\lambda \tau} +\frac{3 \Gamma_{\rm sp} + 2\gamma^{*}}{4 \lambda} \sin{\lambda \tau}\right),
\end{equation}
which depends on the same three parameters $\Omega$, $\Gamma_{\rm sp}$, and $\gamma^{*}$. 

\section{Experimental determination of the QD spontaneous emission rate and dephasing rate}

\subsection{Results from the resonant excitation}

\begin{figure*}[]
\includegraphics[width=150mm]{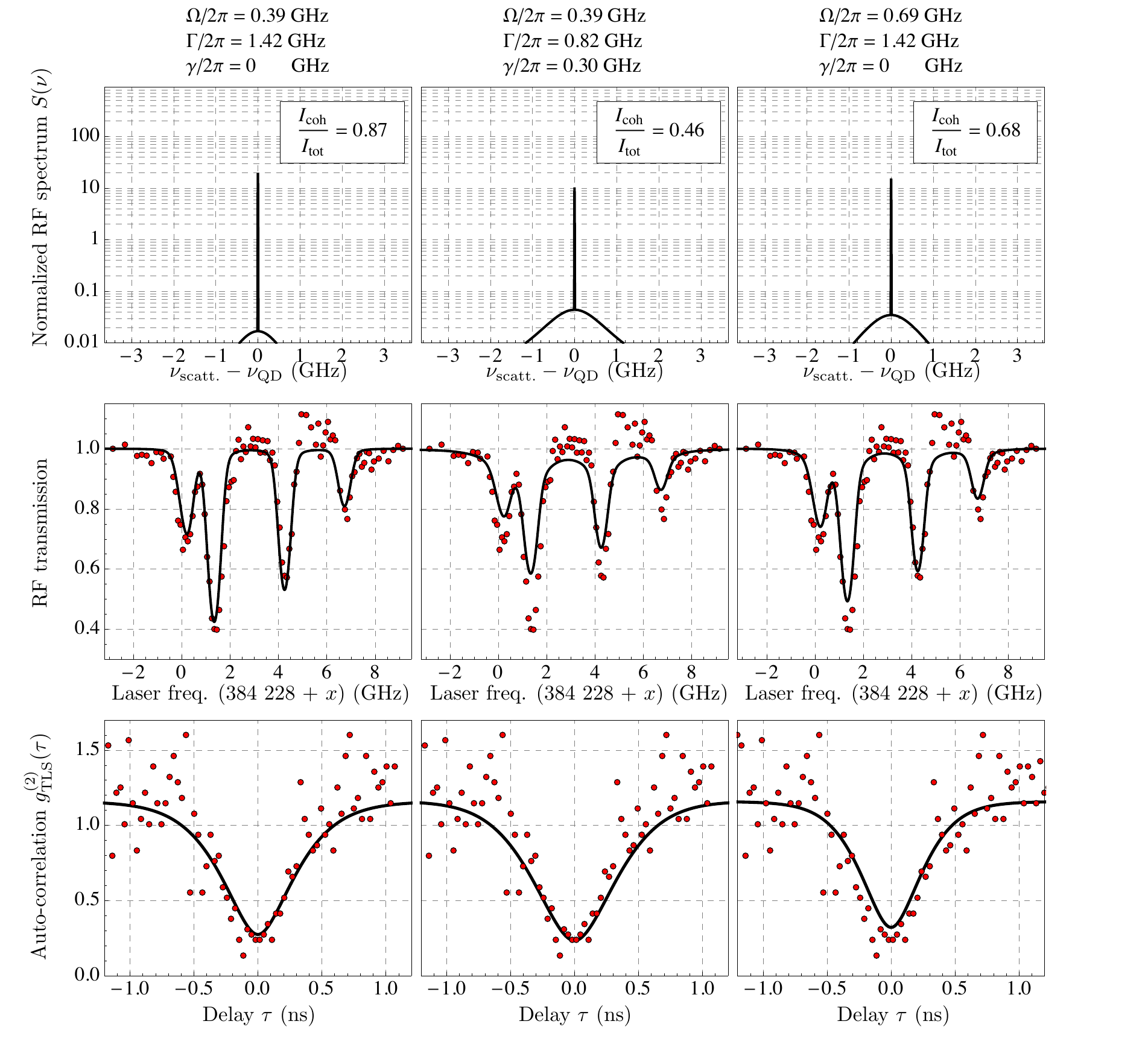}
\caption{{\bf Sensitivity on the fitting parameters.} First row: computed resonant QD spectrum. Second row: absorption spectrum under resonant excitation. Third row: second order correlation function. Open circles correspond to the experimental data.}
\label{Fig:ComparisonPanel}
\end{figure*}

In order to evaluate the QD spontaneous emission rate and dephasing rate, we perform a simultaneous fit ($\chi^2$-minimization) of 1) the Rb vapor transmission spectrum measured with single photons from the resonantly excited QD, and 2) the intensity correlation measurements (respectively Fig.~3b and Fig.~2d). As we used the same resonant laser intensity $I_R=141\,$nW/\textmu m$^2$ in both experiments, the two data sets are fitted by a common set of the three parameters $\Omega$, $\Gamma_{\rm sp}$, and $\gamma^{*}$ (see previous section).
For each data set, the vertical error bars used in the $\chi^2$-minimization result from shot noise in the number of detected photons per time bin. We find $\Omega/2\pi=0.39 \pm 0.10$\,GHz, $\Gamma_{\rm sp}/2\pi=1.42 \pm 0.12$\,GHz, and $\gamma^*/2\pi= 0 \pm \binom{0.10}{0}$\,GHz, where the error bars correspond to one standard deviation.

To appreciate the fit sensitivity, we plot in Fig.~\ref{Fig:ComparisonPanel} the theoretical predictions corresponding to values of the fitting parameters differing by three standard deviations.
\begin{itemize}	
	\item The first column shows the predictions of the model obtained with the parameters from the best fit: $\Omega=2\pi \times 0.39$\,GHz, $\Gamma_{\rm sp}=2\pi \times 1.42$\,GHz, and $\gamma^*=2\pi \times 0$\,GHz.
	
	\item The second column draws attention to the case of a non-zero pure dephasing $\gamma^*=2 \pi \times 0.30$\,GHz (+3\,$\sigma$ value), with the constraint $\Gamma_{\rm sp}+2\gamma^* = 2\pi \times$1.42\,GHz (total FWHM measured in Fig.~2a of the article). In this case, the coherent fraction of the scattered light decreases to $70\%$, such that the absorption peaks on the transmission spectrum become broader and shallower.
	
	\item The last column shows the predictions of the model with larger Rabi frequency $\Omega=2\pi \times 0.69$\,GHz (+3\,$\sigma$-value), keeping $\Gamma_{\rm sp}$ and $\gamma^*$ at optimal values. A close examination shows that the absorption peaks on the transmission spectrum also become broader and shallower, and the rise time at the dip of the intensity auto-correlation becomes slightly shorter.  
\end{itemize}

As an additionnal consistency check, we can fit the dependence of the FWHM of the RF spectrum $\Gamma$ (Fig.~2b) with laser intensity using the expression $\Gamma(I_R)=\sqrt{\Gamma_{\rm sp}^2 + 2 A I_R}$, with the value of $\Gamma_{\rm sp}$ obtained above and an adjustable coefficient $A$. Best agreement is obtained for $ A~=~0.34~\times~10^{17}$\,(rad/s)$^{2}$/(nW/\textmu m$^2$). From this fit, the Rabi frequency corresponding to the operating resonant intensity $I_R=141$\,nW/\textmu m$^2$ that we extract is $\Omega=\sqrt{A I_R}=2\pi \times$0.35\,GHz, in excellent agreement with the value obtained from the previous analysis.

\subsection{Decay-time measurements under non-resonant excitation}
\label{Apdx_NRdecay}

\begin{figure}[]
\includegraphics[width=85mm]{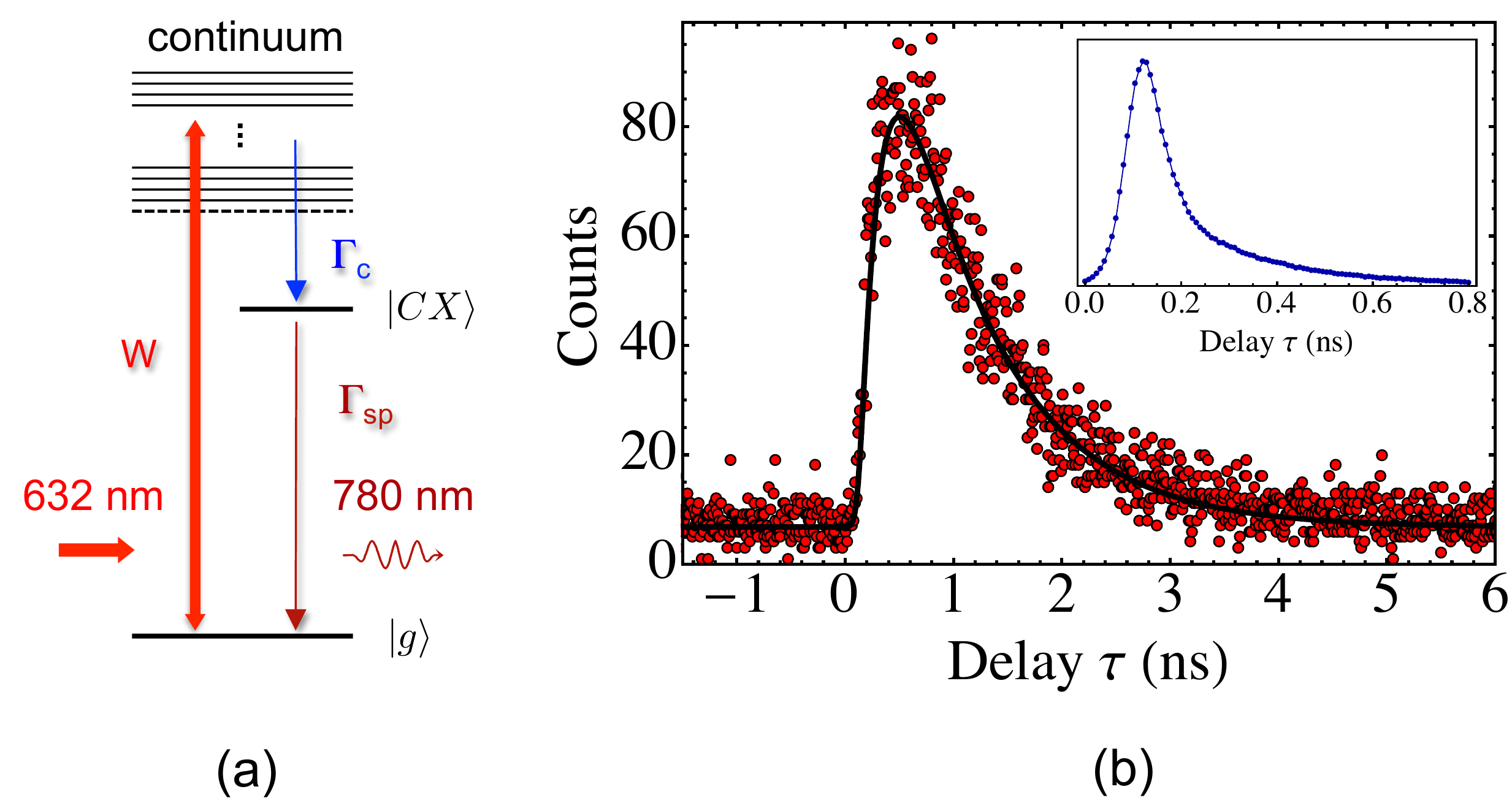}
\caption{{\bf Decay-time measurements.} (a) Non-resonant excitation scheme. (b) Histogram of the QD photons arrival time (8\,ps time bins, integration time 2 minutes). The solid line is a fit using Eq.\ (\ref{Eq:NCXPopulation}) convoluted by the measured instrument response ($ {\rm FWHM} = 100$\,ps, see inset) and scaled to the signal amplitude, with $\Gamma_{\rm sp}=2\pi \times 1.7$\,GHz and $\Gamma_{\rm c}=2\pi \times 176$\,MHz.}
\label{Fig:Lifetime_pulsed_NR}
\end{figure}

In order to confirm the value for the spontaneous emission rate of the QD upper level, we perform decay-time measurements with a non-resonant pulsed laser ($ \lambda =635$ nm, $\simeq 90\,$ps pulses, 80\,MHz repetition rate). The dynamics of the population $N_{\rm CX}(t)$ are well described by Einstein rate equations with two distinct rates: $\Gamma_{\rm c}$, the relaxation from the continuum to the QD excitonic state, and $\Gamma_{\rm sp}$, the radiative decay rate to the QD ground state (see Fig.~\ref{Fig:Lifetime_pulsed_NR}a). Assuming that the system is initially excited in the continuum, the population of the state $\ket{CX}$ takes the form
\begin{equation}
\label{Eq:NCXPopulation}
N_{\rm CX}(t)=\frac{\Gamma_{\rm c}}{\Gamma_{\rm c}-\Gamma_{\rm sp}} \left( e^{-\Gamma_{\rm sp} t} - e^{-\Gamma_{\rm c} t}  \right).
\end{equation}

The result from our measurements is shown in Fig.~\ref{Fig:Lifetime_pulsed_NR}b for low excitation power. The fit of the data yields a high rate (small lifetime) of $2\pi \times (1.7 \pm 0.2)$\,GHz ($\simeq 90$\,ps) and a slow rate (long lifetime) of $2\pi \times (176 \pm 3)$\,MHz ($\simeq 900$\,ps). This is completely consistent with the Rb cell spectroscopy and g$^{(2)}$ results provided the high rate is associated to radiative decay, and the slow rate to relaxation, an association which we have confirmed with pulsed resonant excitation (data not shown). The relationship of the relaxation and decay rates is contrary to the standard interpretation for InGaAs QDs, for which relaxation is much faster than radiative decay. We speculate that the presence of a tunnel barrier between the QDs and the nearby wetting layer (the ring-shaped AlGaAs mound, cf. Fig.~\ref{Fig1}), combined with the indirect bandgap in the Al-rich AlGaAs surrounding matrix are responsible for the unusually slow relaxation dynamics. Of course, the decay curves determines the total decay rate not necessarily the radiative decay rate. However, we are working here with MBE-grown GaAs of very high quality at low temperature where it is safe to assume that non-radiative decay processes are weak such that spontaneous emission represents the dominant decay process.

The radiative lifetime is rather short and corresponds to an oscillator strength of $\sim 100$. The oscillator strength is around 10 in the strong confinement regime \cite{Warburton1998} (quantization energy much larger than the Coulomb energy) rising to well above 100 in the weak confinement regime \cite{Andreani1999}. In this case the result, similar in fact to that of interface fluctuation quantum dots \cite{Guest2002}, shows that the quantum dot is in the intermediate confinement regime.

\section{Complementary information on the blinking in the QD signal}

\subsection{QD1 second order correlation function at long delays}
\label{Apdx_g2QD1long}

\begin{figure}[h]
\includegraphics[width=85mm]{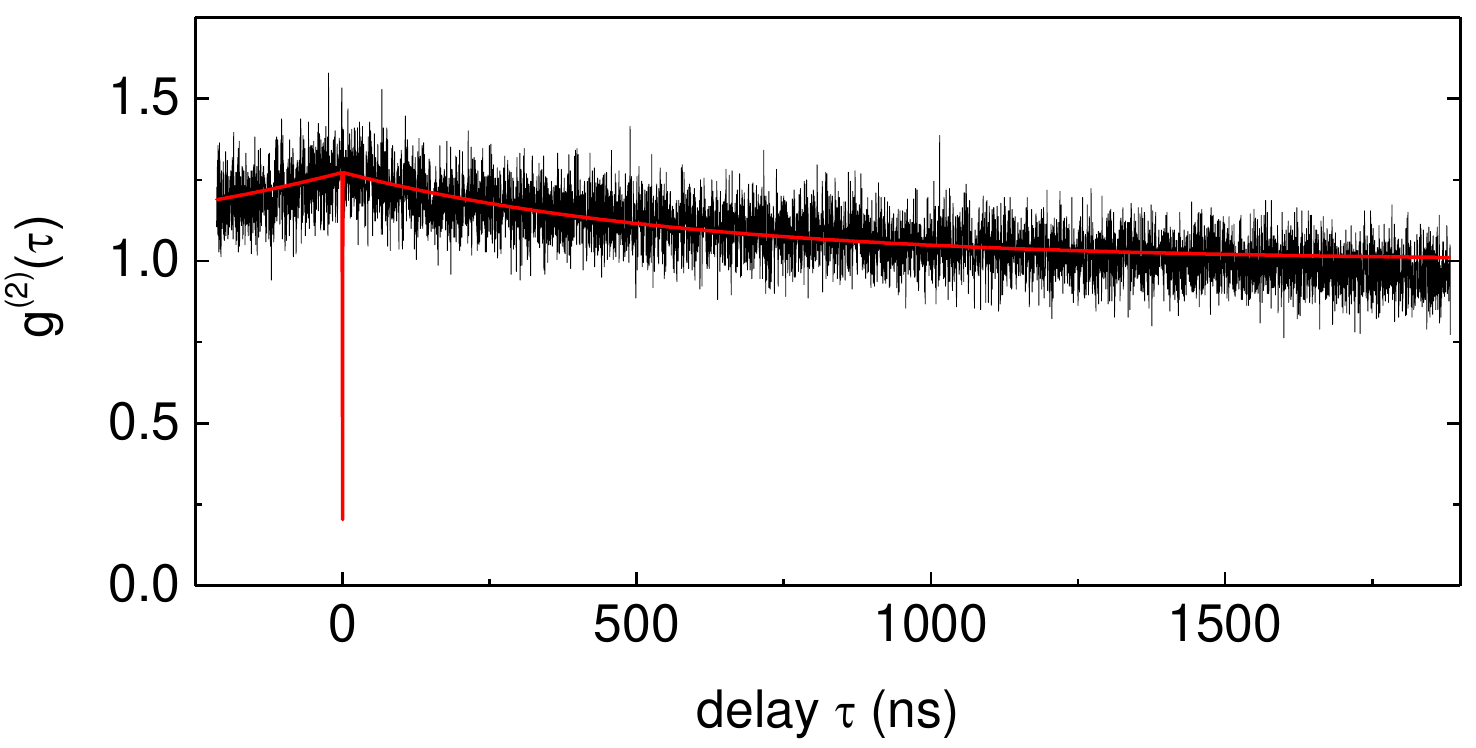}
\caption{{\bf QD1 \boldmath$ g^{(2)}(\tau)$-function at long delays.} Experimental data (black), and a fit (red) using Eq.\ (\ref{Eq:g2blink}), with $\tau_c = 580$ ns and $\beta= 0.8$.}
\label{g2_QD1_longtimes}
\end{figure}

For the correlation measurements, we position a hemispherical solid-immersion lens on the surface of the sample, thereby increasing the count rates by a factor of $\sim 4$.
Fig.~\ref{g2_QD1_longtimes} extends the data shown in Fig.~2d to longer delays. We clearly observe a bunching dynamics with a correlation time  on the order of $600$ ns. The count rate for this experiment was $ 2\times 10^3$~cts/s, so that we can exclude any artefact from the detector\cite{Chopra1972}. 

\subsection{Effect of the non-resonant contribution on the RF signal}

\begin{figure}[h]
\includegraphics[width=85mm]{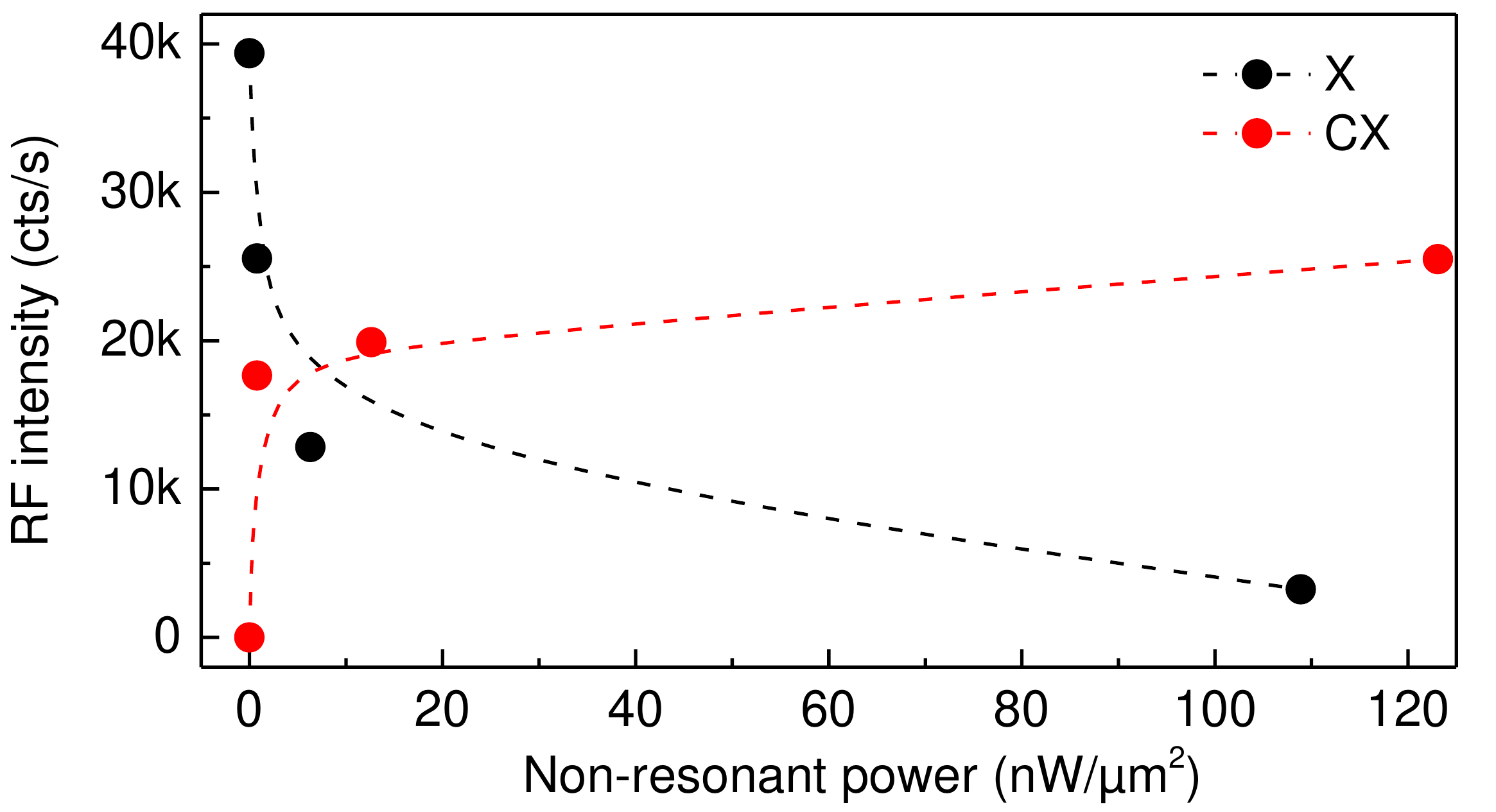}
\caption{{\bf Influence of the non-resonant pump on the RF signal}. Dashed lines are guides to the eyes. }
\label{Suppl-Influence_NR_pump}
\end{figure}

Fig.~\ref{Suppl-Influence_NR_pump} shows the effect of an increasing non-resonant contribution on the resonance fluorescence intensity of the neutral and charged excitons. The data is recorded on QD2 and the data points correspond to the same non-resonant intensities as used in Fig. 4. We note that the values reported here are calculated assuming a perfectly focused beam. Our objective lens is however mono-chromatic and its focus adjusted to maximize collection efficiency at $780$ nm.

%


\end{document}